\documentclass[12pt,preprint,fleqn,5p,twocolumn]{elsarticle}

\pdfoutput=1

\usepackage{graphicx}
\usepackage{amsmath}
\usepackage{amsthm}
\usepackage{amssymb}
\usepackage{sgame}
\usepackage{color}
\usepackage{multirow}
\usepackage{epsfig}
\usepackage{natbib}
\usepackage{setspace}
\usepackage{nccmath}
\usepackage{xfrac}
\usepackage[]{subfig}
\usepackage{accents}
\usepackage{bbm}
\usepackage{algorithm}
\usepackage{algorithmic}
\usepackage{verbatim}
\usepackage{lscape}
\usepackage{afterpage}
\usepackage{caption}
\usepackage{fancyhdr}
\usepackage{fixltx2e}
\fancypagestyle{firststyle}
{
   \fancyhf{}
   \fancyhead[C]{FIRST VERSION DECEMBER 20, 2014. THIS VERSION MARCH 10, 2015.}
}
 \biboptions{super}

\newlength{\dhatheight}

\theoremstyle{definition}

\journal{you.}

\begin{document}

\begin{frontmatter}

\title{Shared intentions and the advance of cumulative culture in hunter-gatherers\tnoteref{t1}}

\author[sa]{Simon D. Angus}
\author[jn]{Jonathan Newton\fnref{fn2}}
\address[sa]{Department of Economics, Monash University.}
\address[jn]{School of Economics, University of Sydney.}
\fntext[fn2]{Come rain or come shine, I can be reached at jonathan.newton@sydney.edu.au, telephone +61293514429. This work was completed while the author was supported by a Discovery Early Career Researcher Award (DE130101768) funded by the Australian Research Council.}
%\tnotetext[fn1]{The authors would like to thank Murali Agastya, Suren Basov, Larry Blume, Zhiwei Cui, David Easley, Sung-Ha Hwang, Claudio Mezzetti, Birendra Rai, Ryoji Sawa, Russell Toth and Galina Zudenkova for detailed comments and fruitful questions, as well as seminar audiences at UNSW, University of Melbourne, Boston University, Cornell University, McGill University, University of Lausanne, University of Zurich, Seoul National University, Sogang University and ANU.}

%\begin{abstract}

%\end{abstract}

\begin{keyword}
Evolution \sep shared intentions \sep  norm \sep technology \sep  social networks.

%\JEL C71 \sep C72 \sep C73 %% or \MSC[2008] code \sep code (2000 is the default)

\end{keyword}

\end{frontmatter}

% \onehalfspacing
\section{Introduction}
 \thispagestyle{firststyle}
It has been hypothesized that the evolution of modern human cognition was catalyzed by the development of jointly intentional modes of behaviour.\cite{vygotsky1980mind,tomasello2014natural,moll2007cooperation} From an early age (1-2 years), human infants outperform apes at tasks that involve collaborative activity.\cite{tomasello2010ape,tomasello2007shared} Specifically, human infants excel at joint action motivated by reasoning of the form ``we will do X'' (shared intentions\cite{tuomela1988,searle1990,bratman1992}), as opposed to reasoning of the form ``I will do X [because he is doing X]'' (individual intentions). The mechanism behind the evolution of shared intentionality is unknown. Here we formally model the evolution of jointly intentional action and show under what conditions it is likely to have emerged in humans. Modelling the interaction of hunter-gatherers as a coordination game\cite{schelling1960strategy}, we find that when the benefits from adopting new technologies or norms are low but positive, the sharing of intentions does not evolve, despite being a mutualistic behaviour\cite{szathmary1995major} that directly benefits all participants. When the benefits from adopting new technologies or norms are high, such as may be the case during a period of rapid environmental change\cite{potts1996evolution,potts2013hominin}, shared intentionality evolves and rapidly becomes dominant in the population. Our results shed new light on the evolution of collaborative behaviours.

\section{Model}
We summarize here the model. Details can be found in Supplementary Information.\cite{AngusNewtonSharedIntentionSupplement} Following Bowles, Choi \& Bowles,\cite{bowles2006group,ChoiBowles2007} consider a metapopulation comprising $m$ partially isolated subpopulations (called demes) of size $n$ individuals. Each individual either has the ability to share intentions (type SI) or does not (type N). Time is divided into generations, each of which comprises $T$ periods. Individuals live for a single generation. Consider a given deme. At the start of a generation, the deme has achieved a level of technological/cultural sophistication $\tau$. This will change as time passes, as will the share of SI and N types in the populations. That is, the model (summarized in Figure 1) is one of gene-culture coevolution. 

\subsection{Within-deme interaction}
The interaction structure within a deme is given by an undirected graph on the set of individuals in the deme. This is determined at the start of each generation, the idea being that for each individual there are a few individuals with whom he has a high degree of interaction (relatives, friends, hunting partners).\cite{ohtsuki2006simple,Apicella:2012dt,Hill:2011bq} At any one time, any given individual plays one of two strategies, the `old' technology, or a `new' technology. For each neighbour who plays the same strategy as he does, he gains a interaction-payoff of either $1$ (old-old) or $\alpha_{\tau} >1$ (new-new). That is, his payoffs in each interaction are given by the coordination game in Table 1. His payoff is then the average of his interaction-payoffs from each neighbour. $\alpha_{\tau}$ represents the within-deme relative fitness benefits of technology $\tau+1$ compared to technology $\tau$.

\subsection{Strategy updating}
%Individuals within any given deme update their strategies together with other individuals as part of a coalition. Any subset of individuals forming a coalition, $S$, must be such that, given the interaction graph, the subgraph induced by $S$ is connected and either $S$ is a singleton or every member of $S$ is of type SI. $k$ is a parameter that gives the maximum number of individuals in any updating coalition. We assume that $k$ is small relative to deme size. Our focal case is $k=2$. At the start of period $t$, randomly choose a coalition $S$ from the set of all coalitions of size $k$ or smaller.
%
Strategies are updated by single individuals but also by pairs of individuals who can share their intentions. A pair of players can only share intentions if both players in the pair are SI types and they are neighbours on the interaction graph. Each period within a generation, either one individual or one pair of individuals is randomly selected to update their strategy. An updating individual or pair of individuals plays a \emph{coalitional better response}, adjusting their strategies so that by doing so they obtain payoffs at least as high as their current payoffs, holding the strategies of all the other individuals fixed.\cite{NewtonAngus2015,Newton1} 
However, when any individual has the opportunity to update his strategy, he will with some small probability make a mistake and switch to a random strategy instead of to his intended strategy.\cite{PY1} 

Note that type (SI or N) does not dictate strategy choice, as it does in traditional evolutionary game theory.\cite{weibull1997evolutionary,dugatkin1998game} Neither does type affect any individual's preferences over profiles of strategies as it does in the literature on evolution of preferences.\cite{guth1998indirect} What type does here is to alter, by enabling or disabling pairwise updating, the set of strategy profiles that can be reached by any given update. Note that the SI type only affects behaviour in the presence of other SI types and that the behaviour manifested by pairs of SI types is mutualistic, in that both participants gain from it (in contrast to altruistic behaviour\cite{szathmary1995major,alvard2002rousseau,smith2003perspectives,rusch2014evolutionary}).

\begin{table}[t]
  \centering
  \begin{tabular}{l c c}
    & \textbf{Old} &\textbf{ New} \\ \hline
\textbf{Old} & $1$ & $0$ \\
\textbf{New} & $0$  & $\alpha_{\tau}$\\\hline
  \end{tabular}
  \caption{Payoffs to within-deme interactions, when the deme has current technology level $\tau$. $\alpha_{\tau} >1$. Entries are interaction-payoffs of an individual whose strategy is given by the row when interacting with an individual whose strategy is given by the column.}
  \label{tab:payoffs}
\end{table}

\subsection{Technology adoption}
Following strategy updating, payoffs for the current period are realized for all individuals. Following this, if the proportion of the individuals in a deme playing `new' is 90\% or higher, we say that the new technology has been adopted, technology $\tau+1$ is now the deme's current technology, and we reset the strategies of every individual in the deme to `old'.

\subsection{Deme extinction}
At the end of each generation, with probability $\eta$, any given deme faces an invader. The invader is another deme chosen at random. If the invader has a higher technology level than the invaded deme, then the invader replaces the invaded deme. The invaded deme becomes extinct and is replaced with a replica (types, current payoffs and tech level) of the invader.
This can represent the possibility of violent conflict between demes, but can equally be considered to model differing extinction and expansion rates of demes with access to different technology. 

\subsection{Reproduction}
Following the extinction phase, each deme reproduces according to a finite population replicator dynamic with mutation rate $\mu$, determining the shares of SI types in the next generation. For simplicity, we assume that reproduction is asexual and haploid. Note that as demes comprise finite numbers of individuals, genetic drift will have an effect in demes that contain both SI and N types.

\begin{figure}[b!]
  \centering
  \scalebox{0.6}
	{
	\includegraphics{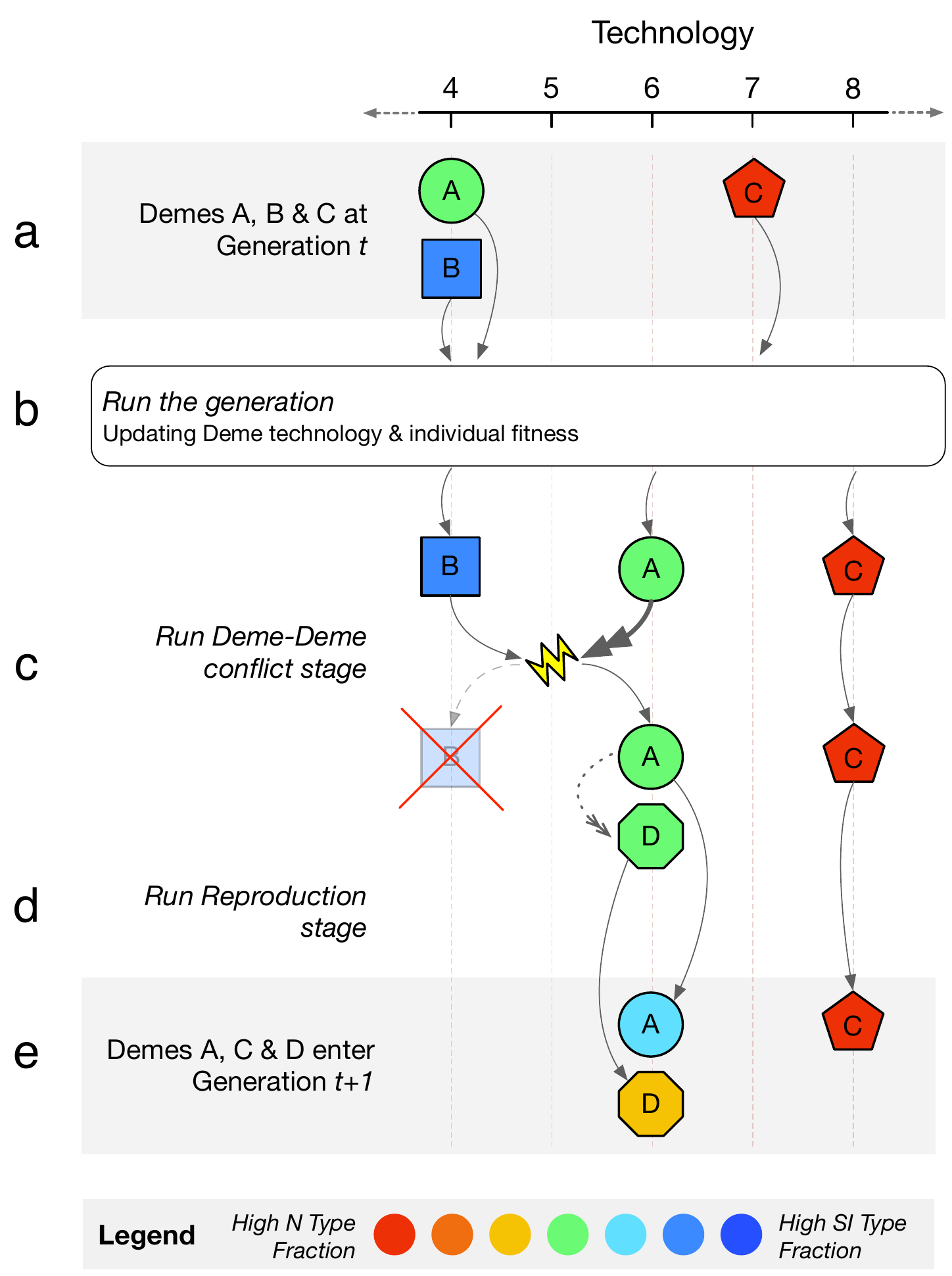}
	}
  \caption{
	\textbf{a:} Demes begin a generation with given technology and number of SI types.
	\textbf{b:} Interaction during a generation gives individual fitnesses and causes advances in deme technology (here, demes A and C increase their tech level).
	\textbf{c:} Some demes (here, deme B) face invasion by other demes (deme A). If the invading deme has higher technology, the invaded deme is eliminated and replaced by a replica of the invading deme (here, deme B is eliminated and replaced by deme D, a replica of deme A).
	\textbf{d:} Demes reproduce and populate the next generation via a finite population replicator dynamic (here, we see within-deme selection and genetic drift in demes A and D changing the number of SI types).
	\textbf{e:} Technology levels and number of SI types are carried forward into the next generation.
	}
  \label{fig:timeline}
\end{figure}

\begin{figure}[b!]
  \centering
  \scalebox{0.7}
	{
	\includegraphics{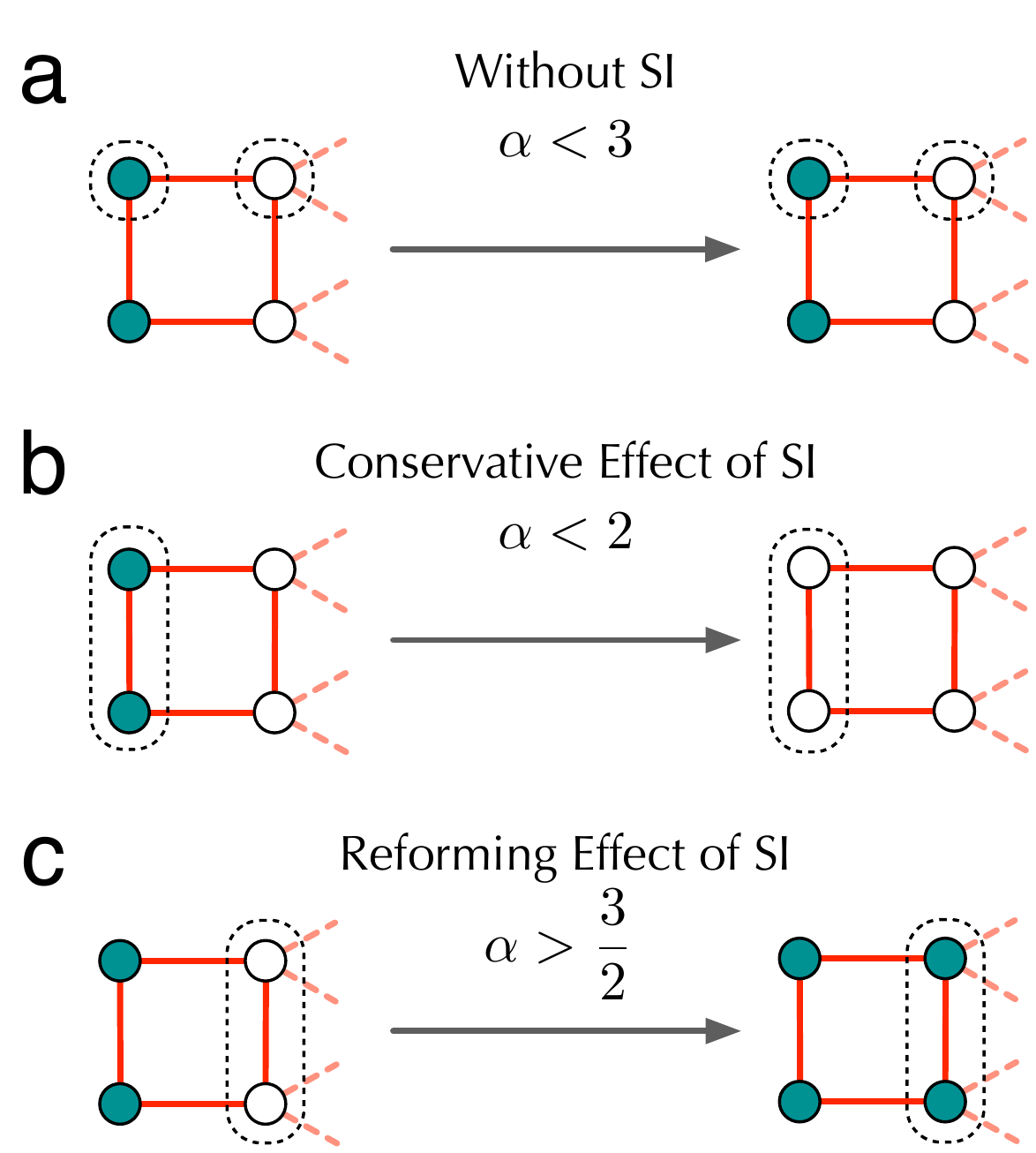}
	}
  \caption{Blue-coloured vertices represent individuals playing `new', white vertices represent individuals playing `old'. Individuals not shown are assumed to be playing `old'. \textbf{a:} In the absence of SI, the only better response for any individual is to retain his current strategy.
\textbf{b:} For low $\alpha$, coalitions of SI type individuals can coordinate payoff improving switches back to `old'.
\textbf{c:} For high $\alpha$, coalitions of SI type individuals can coordinate payoff improving switches to `new'.\protect\cite{NewtonAngus2015}
Note that threshold values of $\alpha$ depend on graph structure and that different interaction structures can yield different thresholds.\protect\cite{NewtonAngus2015,NewtonAngusSpeed}}
 \label{fig:conservativereform}
\end{figure}
%\enlargethispage{-9\baselineskip}

\section{Results}
Relative to demes with low numbers of SI types, demes with high numbers of SI types are slow to adopt new technology when $\alpha_{\tau}$ is low, but fast to adopt new technology when $\alpha_{\tau}$ is high. The former effect arises because, when $\alpha_{\tau}$ is low, SI types playing `new' can coordinate mutually profitable switches back to `old' even when it would not be profitable for any individual acting alone to make such a switch (Figure 2b). Conversely, when $\alpha_{\tau}$ is high, SI types playing `old' can coordinate switches to `new' that would not be profitable for any individual acting alone (Figure 2c).

Hence, if $\alpha_{\tau}$ remains low (conversely, high) over enough generations, then demes with high numbers of SI types will fall behind (conversely, pull ahead) of demes with low numbers of SI types when it comes to technological advancement. 
When SI-poor demes lead in technology, they will outperform SI-rich demes in conflict, and SI will be selected against in the metapopulation. When SI-rich demes lead in technology, the opposite will occur.

When $\alpha_{\tau}$ is low, mutation and genetic drift eventually cause some demes to have low numbers of SI types. These demes gain a technological advantage over other demes, type N becomes dominant and type SI becomes scarce (Figures 3a, 4-Phase I). Conversely, when $\alpha_{\tau}$ is high, mutation, genetic drift, and within deme selection of SI cause some demes to have high numbers of SI types. These demes gain a technological advantage and type SI becomes dominant (Figures 3b, 4-Phase II). These results hold regardless of the initial proportions of SI and N types in the population.
Our results indicate that an extended period of environmental change leading to elevated within-deme fitness benefits from innovation would have sufficed for SI to become widespread.

\section{Discussion}

To the best of the authors' knowledge, this is the first paper to formally model the evolution of collaborative ability, although the work of Bacharach\cite{bacharach2006beyond} makes tentative steps in this direction. Collaboration directly alters \emph{how} strategies are chosen, not \emph{which} strategies are chosen, and although the former affects the latter, how it does so depends on circumstances. The insight we derive, by explicitly modelling techno-cultural advance, is to link technology adoption to the evolution of collaboration. 

As might be expected, large gains from technological adaptation facilitate the evolution of SI.
However, when benefits from new technology are low, collaboration works against a community by slowing its technological advance, even when all members of the community have perfectly aligned interests. Previous literature has discussed how interaction structure can have important implications for cumulative culture.\cite{boyd2002group,Hill:2011bq,hill2014hunter} Our model provides a novel mechanism for this: the social structure within demes combines with the presence or absence of shared intentions and the exogenous technological opportunities of the epoch ($\alpha$) to give varying rates of techno-cultural accumulation. Furthermore, although this study considers the plausible case of scale free social networks\cite{Apicella:2012dt}, there exist a large range of social structures, both regular and random, for which the `conservative' and `reforming' effects of Figure 2 are observed.\cite{NewtonAngus2015,NewtonAngusSpeed} 

The collaborative sharing of intentions is mutualistic, not altruistic: both parties gain from the pairwise adjustment of strategies. Moreover, we look at coordination games, not prisoner's dilemmas. There are no gains to be made from cheating. Despite this, we have seen that the emergence of SI is far from assured.
Thus our model gives a mechanism by which inter-species (e.g. chimp vs. human) differences in benefits from new technologies could lead to diversity in the ability to share intentions. Such differences in the gains from technological advance could arise from physical differences between species, or from differences in environmental variability.\cite{potts1996evolution,potts2013hominin}

\begin{figure*}[t!]
  \centering
  \scalebox{0.7}
	{
	\includegraphics{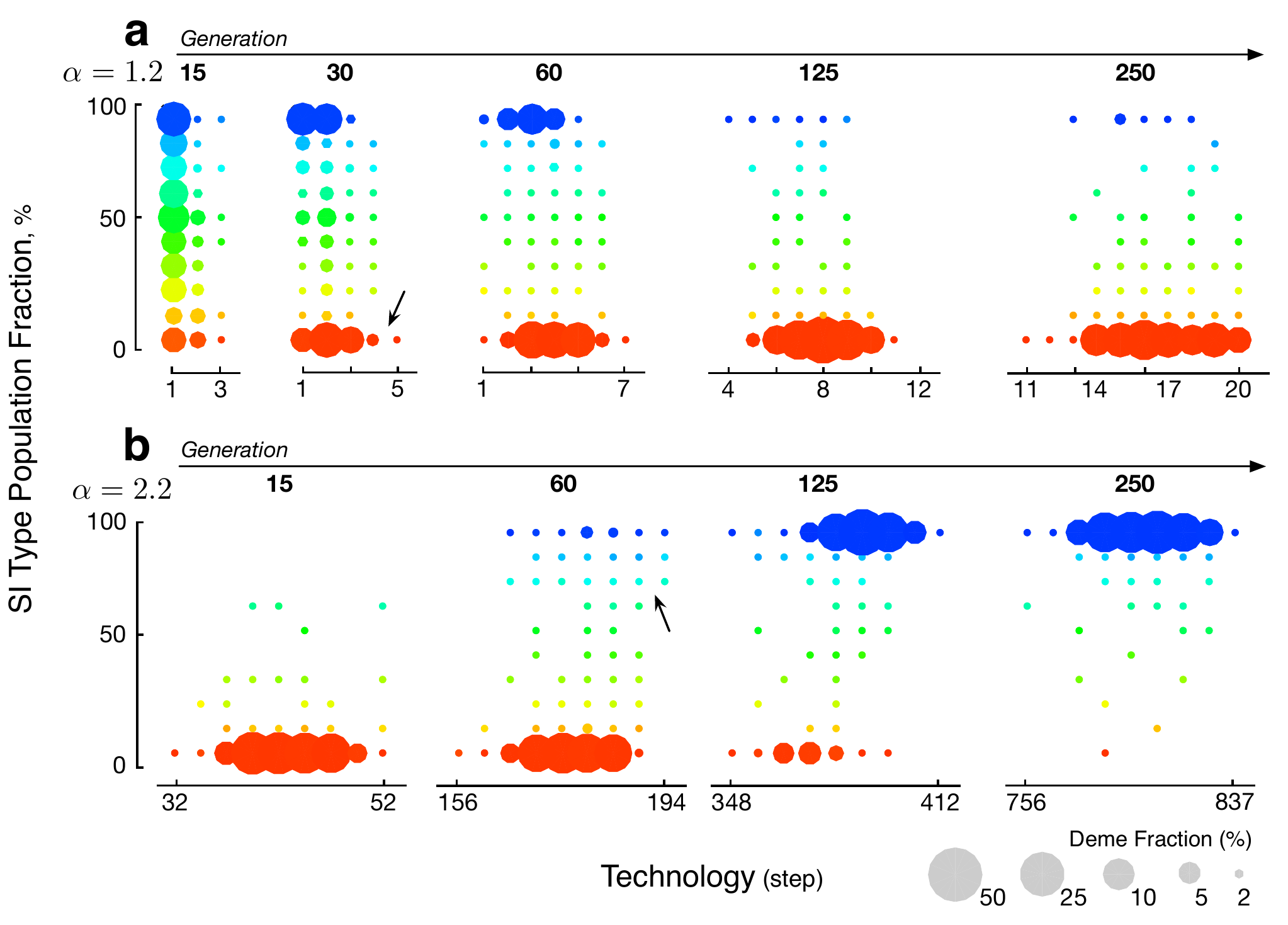}
	}
  \caption{Demes with given fraction of SI type individuals and technology 
level per generation under benchmark conditions, \textbf{a:} $\alpha=1.2$, 
starting from a state in which each individual is SI or N type with equal 
probability,
\textbf{b:} $\alpha=2.2$, starting from a state in which no individuals are SI type. Arrows indicate where demes rich in N and SI types respectively gain a technological advantage.}
  \label{fig:demedynamics}
\end{figure*}

\begin{figure*}[t!]
  \centering
  \scalebox{0.7}
	{
	\includegraphics{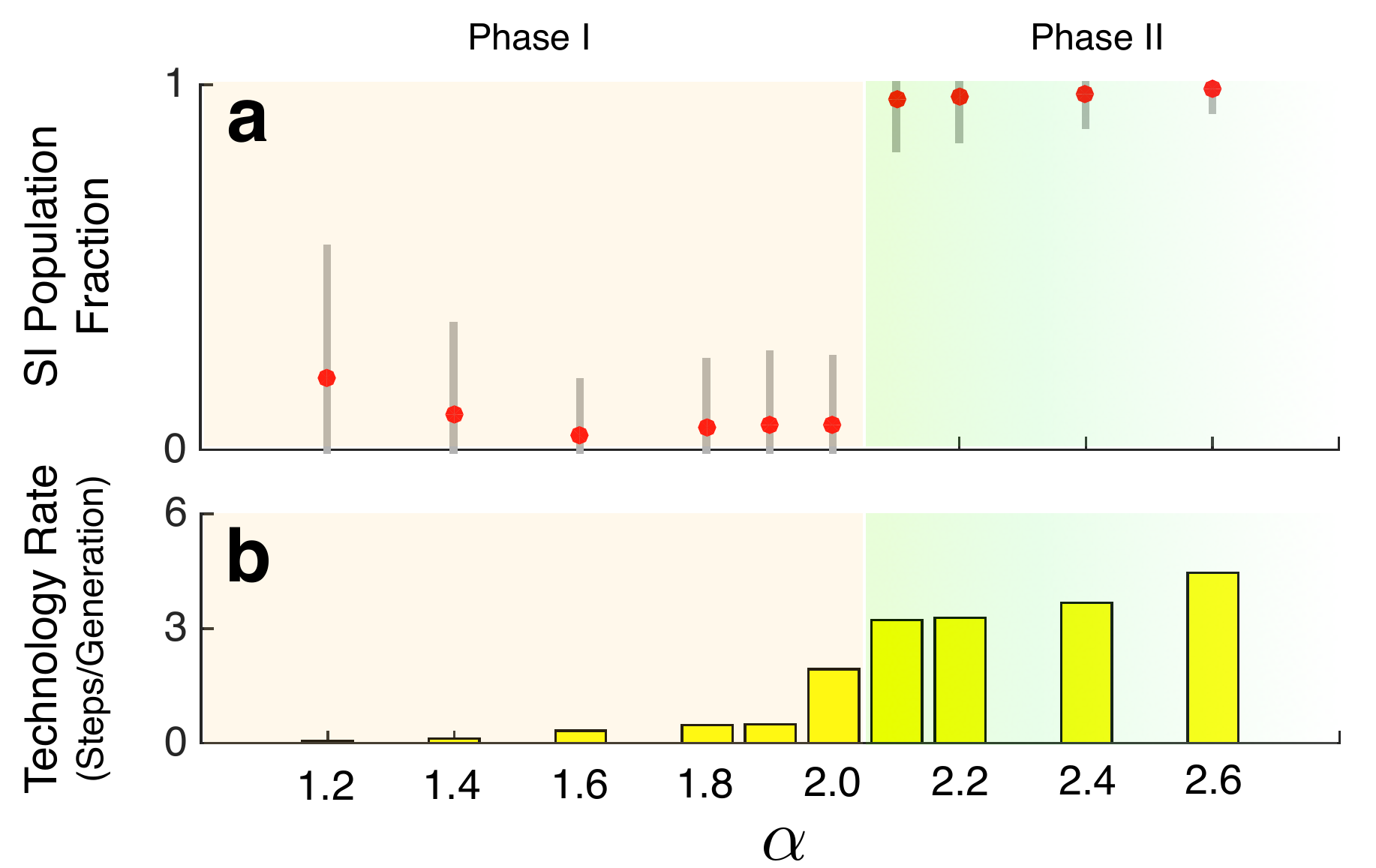}
	}
  \caption{Starting with each individual equiprobably SI or N type under 
benchmark conditions, \textbf{a:} Mean and standard deviation of fraction of SI type individuals across 
all $64$ demes and $10$ replicates during generations 451 to 500,
\textbf{b:} Average rates of technological change (steps per generation) across 
all $64$ demes and $10$ replicates over generations $451$ to $500$.}
  \label{fig:phasetransition}
\end{figure*}
\begin{table*}[b]
\scriptsize
%\begin{minipage}{\textwidth}
%\centering
\begin{tabular}{l l l l}
\hline
\textbf{Determinant} &  & \textbf{Range} & \textbf{Comment/method of estimation} \\
\hline

Number of demes & $m$ & $\mathbf{64}$ & Typical number of languages/dialects in an AIATSIS linguistic zone\cite{AngusNewtonSharedIntentionSupplement}\\
Effective deme size (one-third of census size) & $n$ & $\mathbf{32}$ & Per previous estimates\cite{bowles2006group,Hill:2011bq} \\
Average number of neighbours per individual & $d$ & $4,\mathbf{6},8$ & Degree of scale-free graph (social network)\cite{AngusNewtonSharedIntentionSupplement} \\
Within-deme fitness benefits of new technology & $\alpha$ & $1.2 - 4.0$ & Depends on technology/norm under consideration\\
Periods per generation & $T$ & $\mathbf{2000}$ & $2$ updates per week over $20$ years \\
Maximum coalition size for strategy updating & $k$ & $\mathbf{2}$, $3$, $4$ & Pairwise updating (benchmark) and small coalitions\cite{NewtonAngusSpeed,AngusNewtonSharedIntentionSupplement}\\
Mistake rate in strategy updating & $\varepsilon$ & $0.025,\mathbf{0.05},0.10$ & One mistake every 10-40 updates (benchmark 20) \\
Mutation rate from SI to N and vice versa & $\mu$ & $\mathbf{0.001}$ & For simulation purposes. For lower rates adjust timescales accordingly \\
Per generation conflict probability & $\eta$ & $0.05,\mathbf{0.10},0.15$ & Similar to previous estimates\cite{bowles2006group,AngusNewtonSharedIntentionSupplement} \\
\hline
\end{tabular}
\caption{Parameter estimates. Benchmark values are in bold.}
%\end{minipage}
\end{table*}

%\section*{References}

\def\bibfont{\scriptsize\color[rgb]{0,0,0}}
\bibliographystyle{naturemag}
\bibliography{evolution}

%
%\newpage
%\includepdf[pages=-]{Supplement}

\end{document}